\documentclass[
pra,
reprint,
superscriptaddress,
amsmath,amssymb,
aps,
]{revtex4-2}

\usepackage{times}
\usepackage{graphicx}
\usepackage{dcolumn}
\usepackage{bm}
\usepackage{amsmath,amssymb}
\usepackage{physics}
\usepackage[normalem]{ulem}
\usepackage{url,hyperref}
\usepackage{fixme}
\usepackage{extarrows}
\usepackage{xcolor}

 \fxsetup{status=draft}
 \fxsetup{inline=false, margin=true}
 \fxsetup{theme=color, author=GF}
\DeclareGraphicsExtensions{.png,.jpg,.eps,.pdf}

\begin{document}
	
	\preprint{APS/123-QED}
	
	\title{Overcoming photon blockade in circuit QED single-atom maser with engineered metastability and strong coupling}
	
	\author{A.A. Sokolova}
	\affiliation{Russian Quantum Center, Skolkovo village, Russia}
	\affiliation{Moscow Institute of Physics and Technology, Dolgoprundiy, Russia}
	\affiliation{National University of Science and Technology MISIS, Moscow, Russia}
	\affiliation{Institute of Science and Technology Austria, 3400 Klosterneuburg, Austria}
	
	\author{D.A. Kalacheva}
	\affiliation{Skolkovo Institute of Science and Technology, Moscow, Russia}
	\affiliation{Moscow Institute of Physics and Technology, Dolgoprundiy, Russia}
	\affiliation{National University of Science and Technology MISIS, Moscow, Russia}
    \affiliation{Russian Quantum Center, Skolkovo village, Russia}

	\author{G.P. Fedorov}
	\email{gleb.fedorov@phystech.edu}
	\affiliation{Russian Quantum Center, Skolkovo village, Russia}
	\affiliation{Moscow Institute of Physics and Technology, Dolgoprundiy, Russia}
	\affiliation{National University of Science and Technology MISIS, Moscow, Russia}
	
	\author{O.V. Astafiev}
	\affiliation{Skolkovo Institute of Science and Technology, Moscow, Russia}
	\affiliation{Moscow Institute of Physics and Technology, Dolgoprundiy, Russia}

    \date{\today}
	
	\begin{abstract}
		
		
		Reaching high cavity population with a coherent pump in the strong-coupling regime of a single-atom laser is impossible due to the photon blockade effect. In this work, we experimentally demonstrate that in a single-atom maser based on a transmon strongly coupled to two resonators it is possible to pump over a dozen of photons into the system. The first high-quality resonator plays a role of usual lasing cavity, and the second one presents a controlled dissipation channel, bolstering population inversion, and modifies the energy level structure to lift the blockade. As a confirmation of lasing action, we observe conventional laser features such as the narrowing of emission linewidth and external signal amplification. Additionally, we report unique single-atom features: self-quenching and several lasing thresholds.
		
	\end{abstract}
	
	\maketitle
	
	\section{Introduction}
	Experimental studies on single-atom lasers and masers are following the recent developments in quantum optics of individual quantum systems \cite{haroche2006exploring}. Trapped atoms/ions \cite{firstCs, natureblatt}, superconducting artificial atoms \cite{Astafiev,fluxqubit,Neilinger}, semiconductor double quantum dots \cite{Nomura, Liu2014,thresholddyn}, biased Josephson junctions \cite{Cassidy2017} have already been used as a single-atom gain medium and allowed experimental investigation of non-conventional properties of these lasers such as multistability, emission of squeezed radiation, self-quenching, etc.
    Recently, there was a theoretical proposal \cite{sokolova2021single} regarding a superconducting circuit QED \cite{blais2021circuit} maser consisting of a magnetic-flux-tunable Xmon-type transmon \cite{barends2013coherent} coupled to two microwave resonators: a high-Q reservoir to accumulate microwave photons, and a low-Q auxiliary one. The latter cavity forms an engineered dissipative environment enforcing metastability of the second excited transmon state. Besides bolstering the population inversion, it also significantly modifies the maser energy level structure and allows it to overcome the photon blockade limiting the intensity of the emitted radiation in coherently-pumped systems with strong coupling \cite{blockade, blockade97}. Physical realization of the proposed device would allow one to study those its properties that cannot be explored in numerical simulations, for example the spectrum of the emitted radiation which is hard to compute \cite{Loffer}. Additionally, we expect that suggested architecture will be suitable for novel quantum devices, for example, coherent quantum phonon emitters, which have a wide range of both scientific and practical applications \cite{bolgar2018quantum}. 
	In this paper, we describe our implementation of the suggested architecture and present main results confirming the lasing action of the device.
	
	\begin{figure*}
	\centering\includegraphics[width=\linewidth]{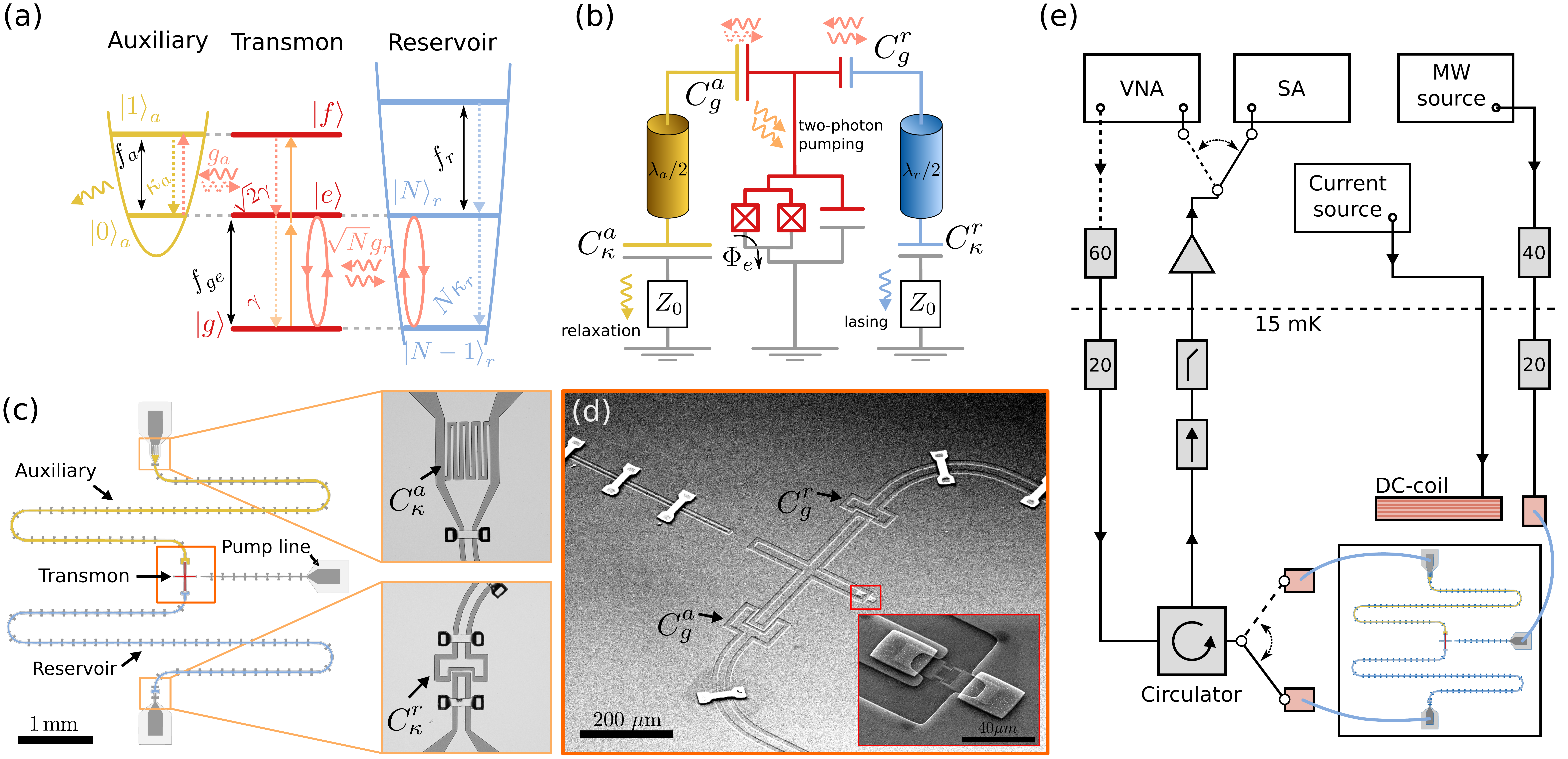}
	\caption{\textbf{(a)} A scheme showing the configuration of the energy levels when the reservoir is resonant with the $\ket{g} \leftrightarrow \ket{e}$ transition, and the auxiliary cavity is resonant with the $\ket{e} \leftrightarrow \ket{f}$ transition. \textbf{(b)} Schematic of the circuit QED implementation of the system. To the left is the auxiliary and to the right is the reservoir cavity, implemented as halfwave transmission line (coplanar waveguide) resonators. Capacitors $C_g^{r,a}$ couple transmon with the reservoir, auxiliary cavity; $C_{\kappa}^{r,a}$ define desired external quality factors. \textbf{(c)} Design of the device (photo-lithography layer) and optical images of the cavity coupling capacitors. Microwave antenna routing the pump signal towards the transmon comes from the right side of the sample. The colors of the elements correspond to the colors on (a) and (b). \textbf{(d)} SEM image of the transmon coupled to the cavities; in the inset, the SQUID area is shown with higher magnification. \textbf{(e)} Scheme of measurement: sample as in (c) is wirebonded, refrigerated to 15 mK, input signals come through attenuated coaxial lines, and output line contains an amplifier chain with a HEMT amplifier. By one switch one chooses whether to measure reflection off auxiliary or reservoir resonator, and by another signal analyzer (SA) and vector network analyzer (VNA) are exchanged; an external solenoid supplies DC magnetic field to the SQUID.}\label{scheme}
	\end{figure*}
	
	\section{Device}
	
	In \autoref{scheme}(a), we reproduce the conceptual schematic of the device from the proposal \cite{sokolova2021single}. The Hamiltonian of the system in the laboratory frame
	\begin{align}
	\label{hamilt}
	  \hat{H} &=  \hat{H}_{t} + \sum_{\lambda=r,a}( \hat{H}^{(\lambda)}_{c} +  \hat{H}^{(\lambda)}_{i}),\\
	  \hat{{H}}_{t} &= h f_{ge}^\text{max}
	  b^{\dagger} b +\frac{h \alpha}{2}
	  b^{\dagger} b( b^{\dagger} b-1),\\
	  \hat{H}^{(\lambda)}_{c}& = h f_{(\lambda)} (a_{\lambda}^\dag a_\lambda + 1/2),\\
	  \hat{H}^{(\lambda)}_{i} &= \hbar g_{\lambda}  ( b a_{\lambda}^\dag + b^\dag  a_\lambda),
	\end{align}
	where $\hat{H}_{t},\ \hat{H}^{(\lambda)}_{c}$ and $\hat{H}^{(\lambda)}_{i}$ are the transmon, cavity and interaction Hamiltonians, where $a_{r,a}$, $b$ are the bosonic annihilation operators for the reservoir, auxiliary resonator and transmon. The cavities are designed to have coupling strengths $g_{r,a}$ and leakage rates $\kappa_{r,a}$. The design values for these parameters have been optimized for maximal steadystate reservoir population $N_\text{ss}$ \cite{sokolova2021single}. The relevant GKSL master equation is based on the collapse operators $\sqrt{\kappa_{r,a}} a_{r,a}, \sqrt{\gamma} b$ for the relaxation of the cavities and the transmon and $\sqrt{\gamma_\phi} b^\dag b$ for transmon pure dephasing. For comparison between the design parameters and the parameters found experimentally in this work, along with the brief description of their meaning, see \autoref{tab:parameters}. The transmon frequency is maximal for zero external flux through its SQUID loop (around 5.95 GHz) and can be tuned down in frequency into resonance consecutively with the reservoir and auxiliary cavity:
	\begin{equation}
	    f_{ge}(\Phi_e) \approx f_{ge}^\text{max} \sqrt{\cos(\pi \Phi_e/\Phi_0)},
	\end{equation}
	where $\Phi_e/\Phi_0$ is the ratio of the external flux through the SQUID to the magnetic flux quantum. There is a slight discrepancy between the anharmonicity of the transmon and the frequency difference between the cavities (180 MHz vs. 145 MHz), and appreciable deviations in $\kappa_a$ and $g_a$ from design values. This mismatch does not critically affect the functioning of the device by virtue of general robustness of the optimal $N_\text{ss}$ to parameter perturbations for large values of $\kappa_a$. Another difference is the non-negligible internal loss of the reservoir cavity which we estimate from the fit to be $\kappa_r^i = \kappa_r - \kappa_r^e = 0.39\ \mu \text{s}^{-1}$ in the worst case (see Supplementary Materials). As $N_\text{ss}$ in the simple model is inversely proportional to total reservoir loss \cite{sokolova2021single}, and as the internal loss in our case approximately equals the external, one can expect halving of the device maximum emission intensity compared to the ideal case. We take $\kappa_r^e$ into account when calculating the $N_\text{ss}$ from measured signal power. Finally, we note that the proper relaxation and coherence times of the transmon were not measured directly due to the limitations of the experimental setup; one can expect its relaxation rate to be lower than $\kappa_r^i$ due to a lower interface participation ratio \cite{wang2015surface}, while the dephasing can be neglected as the system functions in the mixed state \cite{sokolova2021single}.

	\begin{table}
	    \centering
	    \begin{ruledtabular}
	    \begin{tabular}{llll}
	        Parameter & Meaning & Measured & Design \\
	        \hline
	        $f_r$ & reservoir cavity frequency & 5.860 GHz & 6 GHz \\
	        $\kappa_r$ & res. total decay rate & 0.69 $\mu \text{s}^{-1}$ & 0.31 $\mu \text{s}^{-1}$ \\
	        $\kappa_r^e$ & res. decay rate to feedline & 0.3 $\mu \text{s}^{-1}$& 0.31 $\mu \text{s}^{-1}$\\
	        $g_r/2\pi$ & res.-transmon coupling & 11 MHz & 6.5 MHz \\
	        $f_a$ & auxiliary cavity frequency & 5.715 GHz & 5.8 GHz \\
	        $\kappa_a$ & aux. total decay rate & 90 $\mu \text{s}^{-1}$ & 138 $\mu \text{s}^{-1}$\\
	        $g_a/2\pi$ & aux.-transmon coupling & 15.5 MHz & 23.5 MHz \\
	        $f_{ge}^{\text{max}}$ & zero-flux transmon frequency & 5.95 GHz & 6.5 GHz \\
	        $\alpha$ & transmon anharmonicity & 180 MHz & 200 MHz \\
	        $f_r - f_a$ & res.-aux. cavity detuning & 145 MHz & 200 MHz 
	    \end{tabular}
	    \end{ruledtabular}
	    \caption{Summary of the device parameters, measured vs. planned in the design. Associated with dielectric loss, the reservoir total decay rate $\kappa_r$ depends strongly on the measurement power, and the highest value is given here.}
	    \label{tab:parameters}
	\end{table}
	
	The electrical scheme of the device and the design for nanofabrication are shown in \autoref{scheme}(b,c). The couplings and dissipation rates are physically determined by the capacitances $C_{\kappa, g}^{a,r}$ which are small enough to be produced in a single Al layer. However, as they break the ground plane, we also use air-bridges to ensure the uniformity of ground electric potential. In \autoref{scheme}(d) we show a SEM image of the Xmon along with its SQUID in the inset (see Supplementary Materials for fabrication details). 
	
	\section{Reflection and emission spectroscopy}
	
\begin{figure}
		\centering\includegraphics[width=0.49\textwidth]{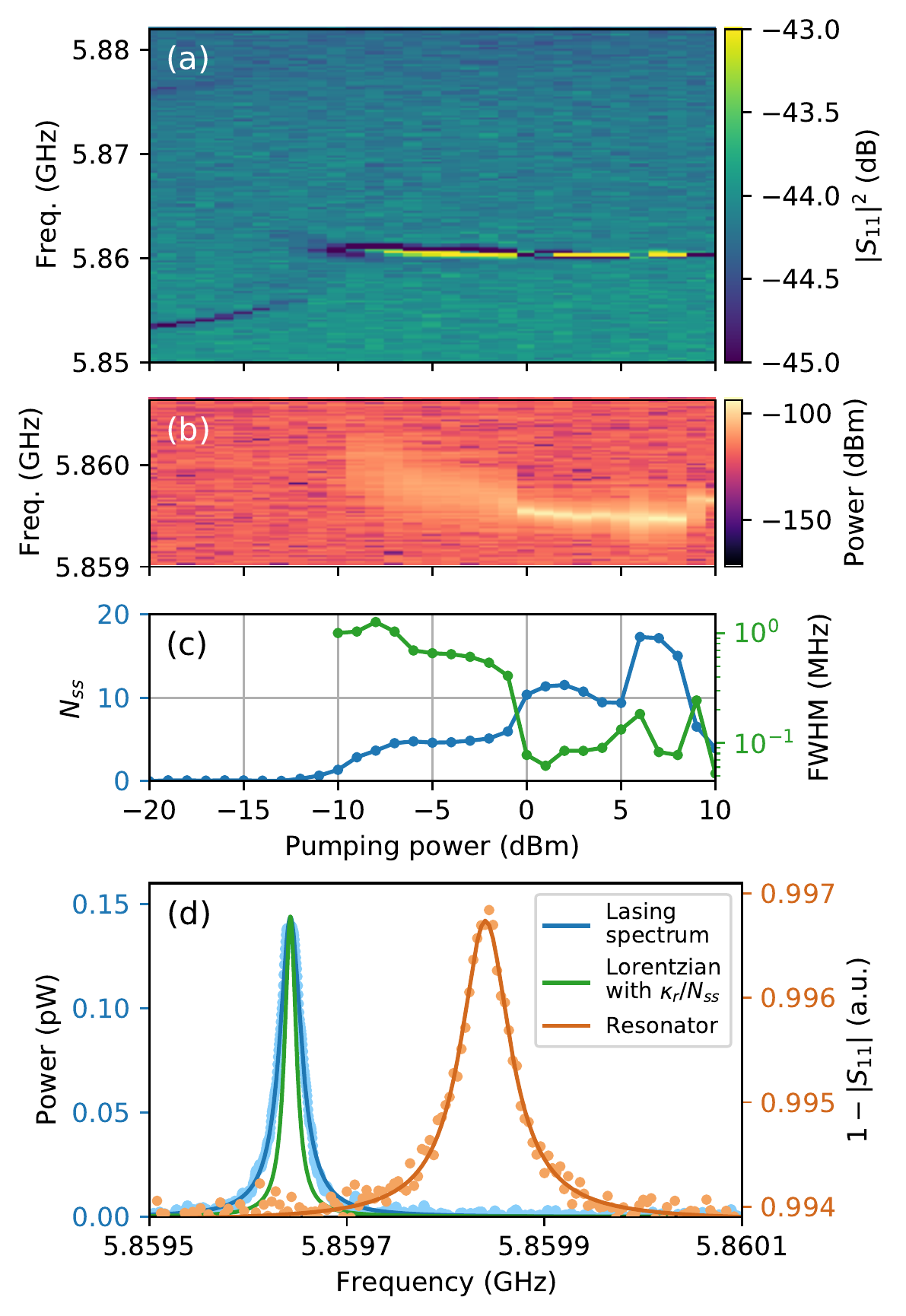}
		\caption{\textbf{(a)} Reflection spectroscopy of the transmon-resonator avoided crossing vs. the two-photon pumping power, pumping frequency $f_p = 5.788$ GHz, VNA signal power is at -50 dBm. \textbf{(b)} Measured emission spectrum from the reservoir cavity vs. the pumping power. \textbf{(c)} Estimated number of photons in the reservoir $N_\text{ss}$ and the spectral linewidth calculated from (b) by Lorentzian fitting. \textbf{(d)} Lasing spectrum (blue, FWHM: 26 kHz) for 10 dBm pumping power compared with the resonator reflection profile for the same photon population ($N_\text{ss} = 3.9$, orange, 61 kHz) and the Schawlow-Townes limit (green, 16 kHz).}\label{main_lasing}
	\end{figure}

	We begin the study of the system by performing reflection spectroscopy of the reservoir resonator when the transmon is tuned into resonance with it, see \autoref{main_lasing}(a). The vector network analyzer (VNA) measuring the complex reflection parameter $S_{11}$ is set at a low power near the single-photon regime for the reservoir cavity (frequency sweep range of the VNA is shown on the $y$-axis). While performing spectroscopy, we send an additional signal at $f_p = f_{gf}/2 = 5.788$ GHz to the pump line using a separate microwave source. Its power is shown on the $x$-axis. When its power is off or very low, two transitions of the avoided crossing $\ket{0_a,\, g,\, 0_r} \rightarrow (\ket{0_a,\, g,\, 1_r} \pm \ket{0_a,\, e,\, 0_r})/\sqrt{2}$ are visible at 5.854 and 5.876 GHz, from what $g_r/2\pi=11$ MHz (see Supplementary Materials). When the two-photon pumping power is increased, these transitions become Stark-shifted upwards and saturate. They completely disappear at around -10 dBm level of the microwave source, which is the beginning of the lasing action. The $|S_{11}|$ characteristic of the remaining spectral line is not a Lorentzian dip above this power but shows small amplification -- manifestation of the injection locking effect \cite{injection_locking}. Also, above -10 dBm the reservoir begins to emit power (\autoref{main_lasing}(b)) which can be detected upon disconnecting the VNA from the measurement setup and connecting a signal analyzer (SA) to the output line. 
	
	Precise calibration of the emission power is difficult due to the uncertainty of the amplification level in the output line. We find it by using the fact that the vacuum Rabi splitting can be regarded as an effective two-level system, for which calibration of absolute input power is feasible \cite{bishop2009nonlinear, honigl2020two}. First, we measure the saturation of the avoided crossing peaks with increase of VNA power $P_\text{VNA}$ and compare the data with numerical simulation for the extracted system parameters; this allows us to find the driving amplitude, and thus the absolute signal power $P_\text{c.p.}$ at the reservoir coupling port for a given $P_\text{VNA}$. Next, we measure the power of a slightly off-resonant reflected and amplified signal $P_\text{SA}$ by the SA assuming almost unity reflection off the coupling port in the vicinity of the resonance. Finally, calculate the amplification level $a = P_\text{SA}/P_\text{c.p.}$ in the output line (see details in the Supplementary Materials). When the amplification is known, the total emission power at the coupling port can be calculated as an integral of the signal power spectral density $S_{VV}(f)$ measured by the signal analyzer as
	\begin{equation}
	P = \kappa_r^e N_\text{ss} \hbar\omega = \frac{1}{a}\int_{\text{BW}} S_{VV}(f)\, df,  
	\end{equation}
	where BW stands for the bandwidth where the emission signal is non-negligible. For this value, we take the span of the SA shown in the y-axis of \autoref{main_lasing}(b). We plot the calibrated $N_\text{ss}$ vs. pumping power in \autoref{main_lasing}(c) and confirm that it is indeed possible to achieve significant reservoir population using our architecture: in the range between 5 and 10 dBm of pumping power, reservoir accumulates nearly 20 photons. According to the model described in the proposal and generalized to the case of non-zero detuning between the transitions $g\rightarrow e$ and $\ket{0}_a \rightarrow\ket{1}_a$, for experimental parameters the system should accumulate around 24 photons, which is in a good agreement with data. Using a numerical model with the experimental parameters, we find even better agreement for maximal $N_\text{ss}$ around 19 (for details on both approaches, see Supplementary Materials).
	
	In \autoref{main_lasing}(c), one can observe four lasing thresholds in $N_\text{ss}$ located at -10, -1, 5 and 8 dBm which are accompanied by changes in the emission linewidth and lasing frequency. This behavior is probably connected with the special energy level structure of the system \cite{sokolova2021single} and can be qualitatively reproduced in simulations (see Supplementary Materials). After the last threshold, there is a significant decrease of $N_\text{ss}$, which we identify as the self-quenching effect \cite{oneatomlasers}.
	
	Finally, in \autoref{main_lasing}(d) we compare the lasing spectrum and reservoir cavity response, which both can be well approximated by Lorentzian curves. The orange line shows $1-|S_{11}|$ when the transmon is far detuned, and the blue line is the lasing spectrum at 10 dBm pumping power. The central frequency of the latter is 200 kHz lower than $f_r$, which may be explained by the fact that frequencies of the transitions between high energy levels are not equal to $f_r$ due to strong coupling \cite{sokolova2021single}. For high reservoir population, the line of the emission spectrum is significantly narrower than the resonator proper amplitude characteristic: the FWHM is 26 kHz vs. 61 kHz for $N_\text{ss}=3.9$, and approaches the Schawlow-Townes-limited spectrum with a width of $\kappa_{r}/2\pi N_\text{ss} = 16$ kHz.
	
\begin{figure}
	    \vspace{0.8cm}
		\centering\includegraphics[width=0.45\textwidth]{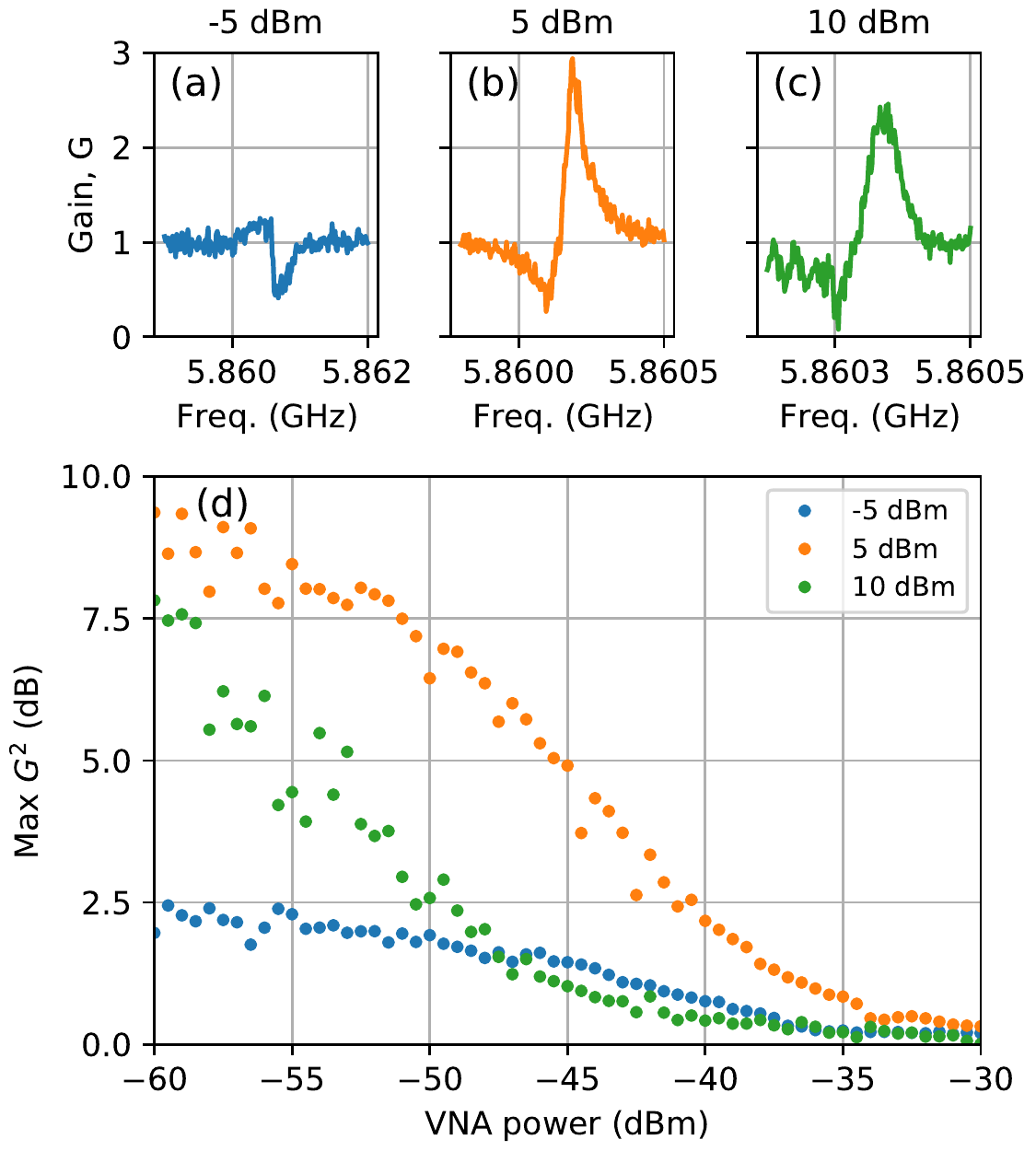}
		\caption{\textbf{(a-c)} Gain of reflection amplitude vs. frequency for pumping powers -5, 5 and 10 dBm, respectively, at the VNA power of -60 dBm. \textbf{(d)} Maximum value of the power gain over the frequency range shown on the 
		$x$-axes in (a-c) depending on the VNA power for three pumping powers. }\label{amplification}
	\end{figure}

	As an additional test for the laser action of the device, we show that an external microwave signal is amplified upon reflection off the reservoir cavity when the system is pumped above the first threshold. The results are presented in \autoref{amplification} for three different pumping powers: -5 dBm (a), 5 dBm (b) and 10 dBm (c). The reference level is depicted by a horizontal black line, and the dip in the data trace means resonant absorption of radiation, while the peak reveals amplification. Amplification depends non-linearly on the pumping power, with a maximum amplitude gain of $G=3.6$ at 5 dBm pumping power. In \autoref{amplification}(d), one can see how the amplification saturates with increased VNA power for the three values of pumping power, which is a typical behavior \cite{injection_locking}. We also observe that in \autoref{amplification}(a) the amplification area is located slightly below resonance, while in \autoref{amplification}(b-c) it is slightly above. Based on the data shown in \autoref{main_lasing}(a), we find that the transition between these regimes occurs at 0 dBm of pumping power, and it coincides with the second lasing threshold in \autoref{main_lasing}(c); the nature of this effect remains unclear to us.

	\begin{figure}
		\centering\includegraphics[width=0.45\textwidth]{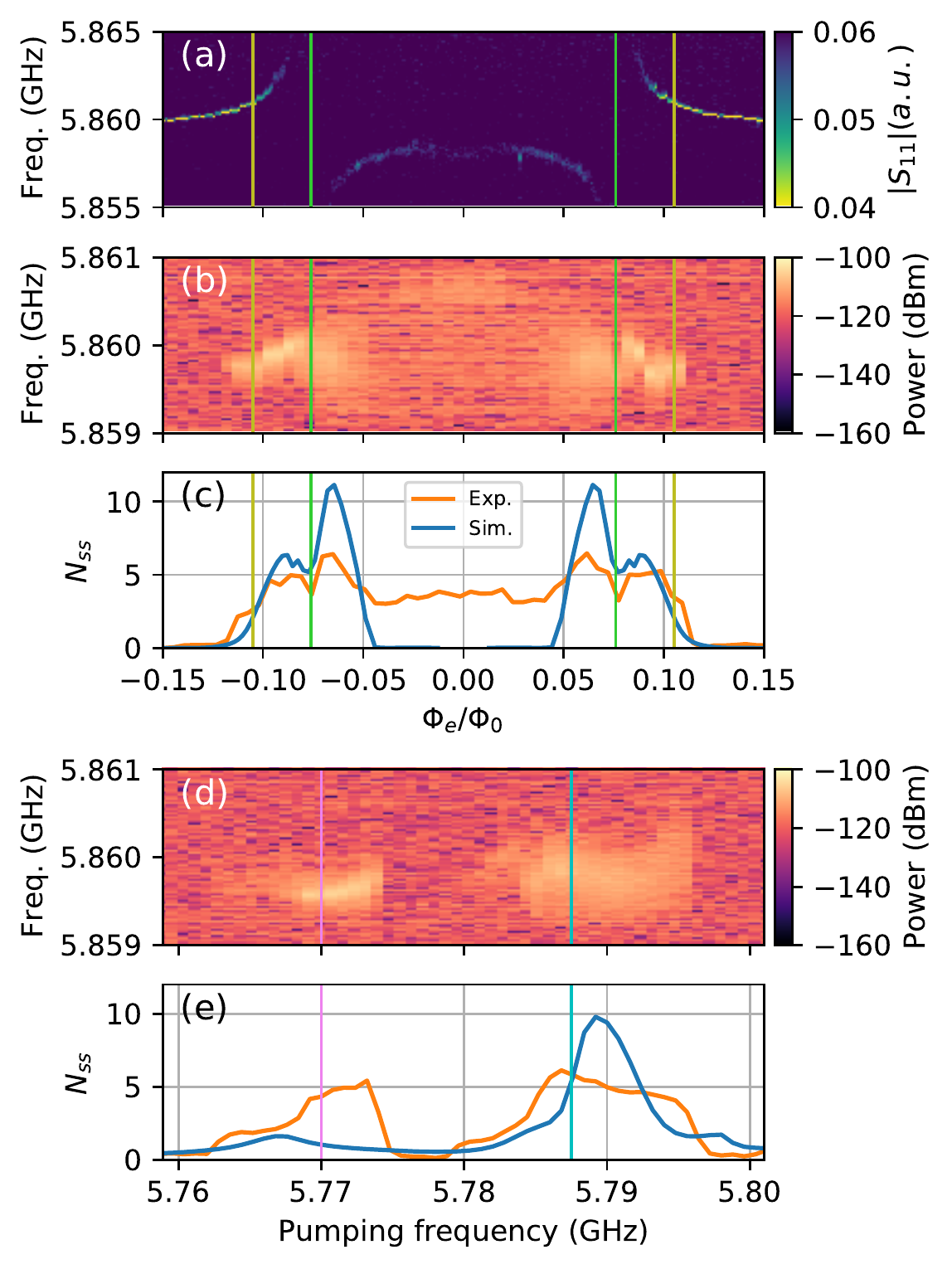}
		\caption{\textbf{(a)} Reflection spectroscopy of the avoided crossing at -60 dBm on the VNA vs. magnetic flux $\Phi_e$ through the transmon SQUID. Green lines mark the points of transmon-reservoir resonance and yellow lines show $\Phi_e$ where the $ge$ frequency is equal to the pumping frequency. \textbf{(b)} Lasing spectrum vs. $\Phi_e$ for pumping power of -5 dBm at 5.788 GHz. \textbf{(c)} Reservoir $N_\text{ss}$, experiment (orange) and simulation (blue) vs. $\Phi_e$. \textbf{(d)} Lasing spectrum depending on the pump frequency when its power is -5 dBm. Magenta line is the two-photon frequency of $\ket{g} \leftrightarrow \ket{1}_a$ process, cyan line~-- of $\ket{g} \leftrightarrow \ket{f}$ process. \textbf{(e)} Number of photons in the reservoir resonator vs. pumping power, experiment (orange) and simulation (blue).}\label{flux_freq}
	\end{figure}

	In order to quantify the device action outside the optimal regime and better understand the lasing mechanism, we also study the dependence of the emission spectrum on the external flux in the SQUID and the pumping frequency. We now use lower-power pumping (-5 dBm) to avoid power broadening and achieve better resolution of transitions. \autoref{flux_freq}(a) shows scan of two symmetric avoided crossings with green lines marking the points where the transmon $ge$ transition is resonant with the reservoir, and yellow lines~-- where the $ge$ transition is resonant with the pump signal. The first configuration was studied before in \autoref{scheme}(a), and in the second one the qubit may be pumped directly. As can be seen in \autoref{flux_freq}(b) and (c), the reservoir population is the highest in proximity to the green lines, both in the experiment and the simulation. Nevertheless, we note that emission is observed for a wider frequency range in the experiment than in the simulation. A small dip in $N_\text{ss}$ which coincides with the central points of the avoided crossings can be explained by the fact that low-power pumping cannot overcome the photon blockade, being under the lasing threshold \cite{sokolova2021single}.
	
	Similarly, we measure the dependence of the emission spectrum on the pumping frequency for the resonant configuration. The results are shown in \autoref{flux_freq}(d). Since there is a 35 MHz detuning between $f_r + f_a$ and $f_{gf}$ when $f_{ge}=f_r$, we mark two competing cases of two-photon pumping: the magenta vertical line corresponds to the resonance with the transition from the ground state to $\ket{1_a,\, e,\, 0_r}$ (this is the configuration chosen in \autoref{main_lasing}), and the cyan one -- to $\ket{0_a, \,f, \, 0_r}$. In \autoref{flux_freq}(e), two areas of high emission attributed to these processes are clearly visible. We find that in contrast with the simulation, in the experiment both pumping frequencies give comparable values for the integrated emission power and the corresponding photon number in the reservoir; however, pumping at $f_{gf}/2$ gives a wider emission line than at $(f_r + f_a)/2$. We also note the asymmetry of the emission areas with respect to the corresponding marker lines, showing a notable shift towards higher pump frequencies.

	\section{Discussion}
	
	In this work, we have implemented a single-atom maser based on a transmon that overcomes the photon blockade in the strong coupling regime, concluding previous theoretical study on the subject. We were able to pump more then 15 photons in cavity, a notably high population for a coherent pump in this class of systems \cite{natureblatt, blockade, blockade97, sokolova2021single}. Additionally, we observed several lasing thresholds and self-quenching accompanied by variations in emission spectrum central frequency and linewidth despite that usually single-atom lasers in the strong-coupling regime are thresholdless \cite{comment, natureblatt, firstCs}. We explain the possibility of overcoming the photon blockade in our device by the additional splittings in the level structure emerging from coupling to auxiliary resonator and ensuring resonance condition for the two-photon pump for high populations. Besides, based on theoretical studies we assume that our system should exhibit such features as bistability of Wigner function and sub-Poissonian statistics of emitted radiation \cite{sokolova2021single}.
	
	As this was the first experimental study of this kind of a device, we have also investigated in more detail the conventional manifestations of the lasing effect to confirm its nature. Firstly, we have demonstrated that the emission linewidth is narrower then the resonator proper linewidth and, notably, approaches the Schawlow-Townes limit for certain regimes. Secondly, if supplied, an external microwave signal is being amplified by the device; this effect was previously demonstrated for single-atom masers \cite{Astafiev} as well as in ordinary lasers and is caused by injection-locking effect \cite{injection_locking}. Finally, we measured the emission spectrum depending on transmon frequency and pumping frequency and observed that lasing manifests itself when the system parameters are close to the theoretically predicted, which is a strong evidence of the correctness of our model.

	\section{Acknowledgments}\label{app:acknowledgments}
    We thank N.N. Abramov for assistance with the experimental setup. The sample was fabricated using equipment of MIPT Shared Facilities Center. This research was supported by Russian Science Foundation, grant no. 21-72-30026.

	\bibliography{main}

\end{document}


\title{Supplementary material for "Overcoming photon blockade in circuit QED single-atom maser with engineered metastability and strong coupling"}

	
	\maketitle

	\section{General measurement notes}\label{app:measurement}
	
	Our experiments were done in two cooldowns of the dilution refrigerator. Single-tone and two-tone spectroscopy for device characterization (including \autoref{app_qubit}, \ref{app_ph} and  Fig. 4(a) of the main text) and auxiliary resonator characterization were done in the first cooldown, and all the other measurements -- during the second cooldown. In the second cooldown the attenuation in the input line was increased by 15 dB.
	
	\section{Device parameters}\label{app:device}
	
	\subsection{Reservoir resonator}

\begin{figure}[b]
		\centering\includegraphics[width=0.6\textwidth]{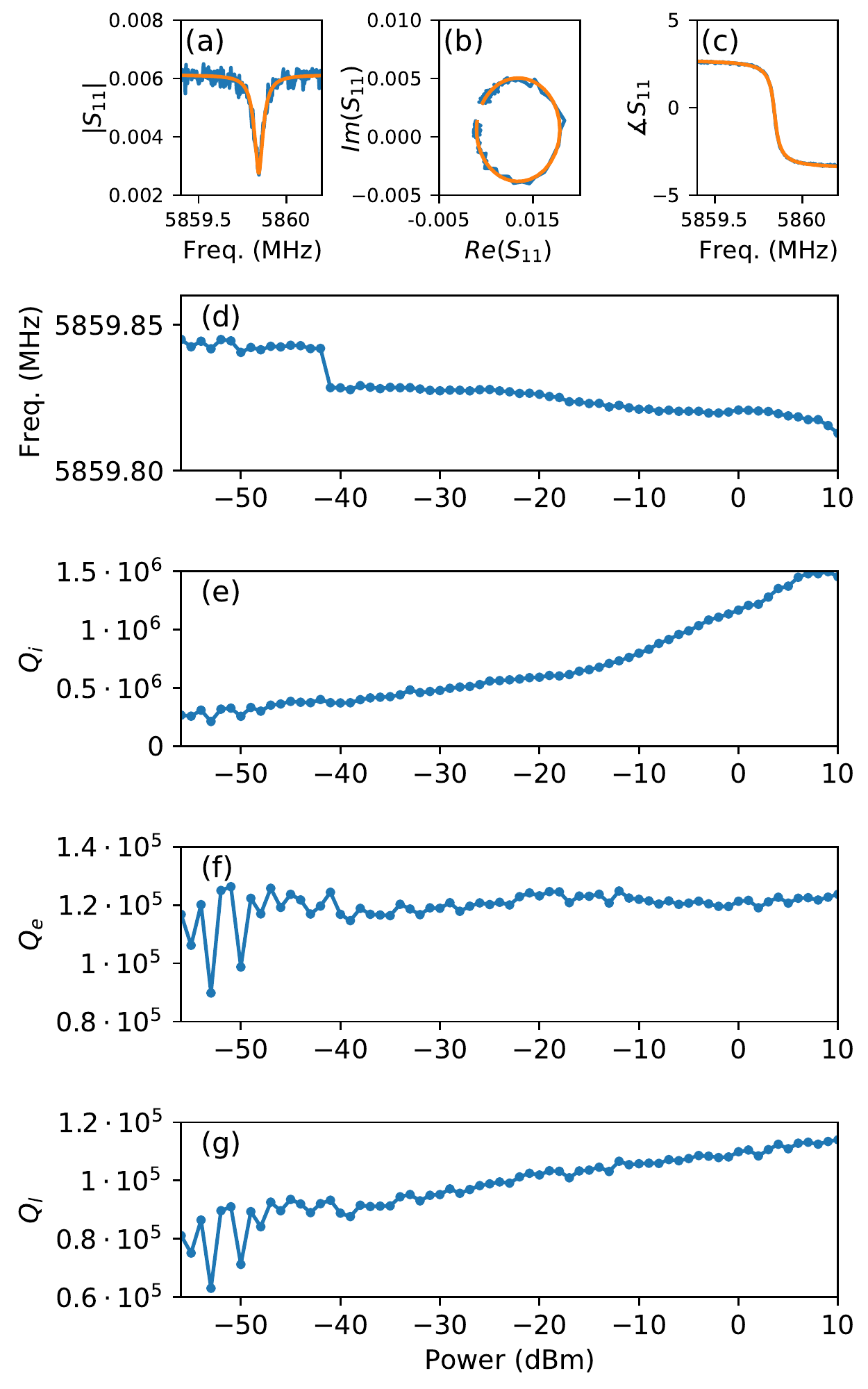}
		\caption{\textbf{(a-c)} Fit of the reservoir resonator (VNA power at -50 dBm) \textbf{(d-g)} Frequency, internal Q-factor ($Q_i$), external Q-factor ($Q_{e}$) and loaded Q-factor ($Q_{l}$), respectively, depending on power.}\label{app_res}
	\end{figure}

	The parameters of the reservoir resonator were measured using VNA when the qubit was far detuned from the resonator. Complex VNA traces were fitted using Python package \textit{resonator\_tools} \cite{Probst} and shown in \autoref{app_res}(a-c); extracted values for the resonance frequency and the quality factors are shown in \autoref{app_res}(d-g).
	
	Afterwards, reservoir frequency was assumed to be $f_{r} = 5.860$ GHz in numerical simulations and all other calculations. The dissipation parameters were calculated from the fitted Q-factors: $\kappa_{r} = 0.69\ \mu\text{s}^{-1}$, $\kappa^{e}_{r} = 0.3\ \mu\text{s}^{-1}$. The quality factors changed slightly between the cooldowns, but qualitatively it did not affect lasing properties (see \autoref{app_sim}).

	\subsection{Auxiliary resonator} 
	
\begin{figure}[b]
		\centering\includegraphics[width=.8\textwidth]{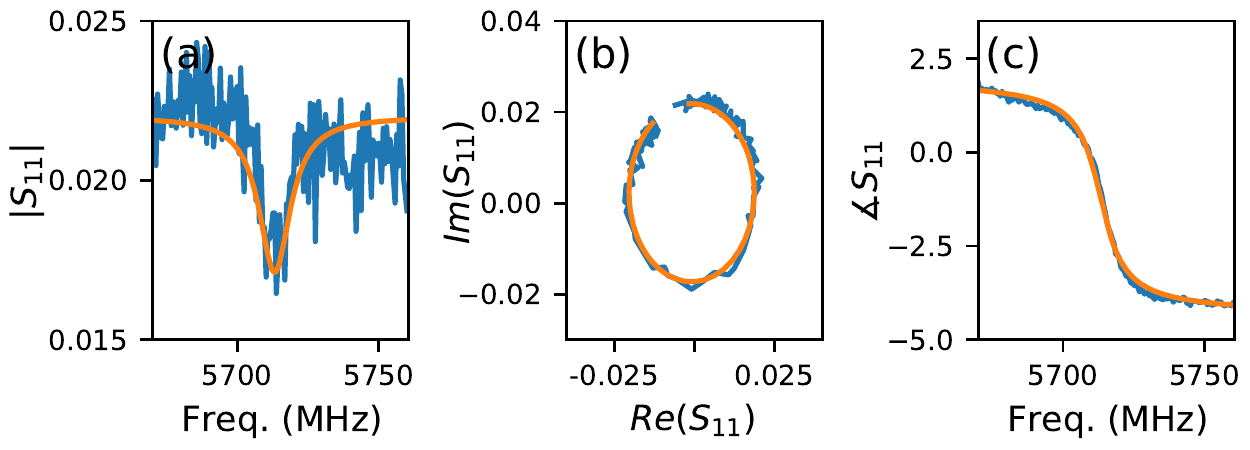}
		\caption{\textbf{(a-c)} Manual fit of the auxiliary resonator (VNA power at -50 dBm)}\label{app_aux}
	\end{figure}

	Automatic fitting of auxiliary resonator was not feasible difficult due to its low Q-factor; the data was fitted manually (see \autoref{app_aux}) and the loaded Q-factor was estimated to be around 400. The following values were used for simulations: $f_{a} = 5.715$ GHz, $\kappa_{a} = 90\ \mu\text{s}^{-1}$.
	
	\subsection{Transmon and coupling} 
	
\begin{figure}[h]
		\centering\includegraphics[width=.8\textwidth]{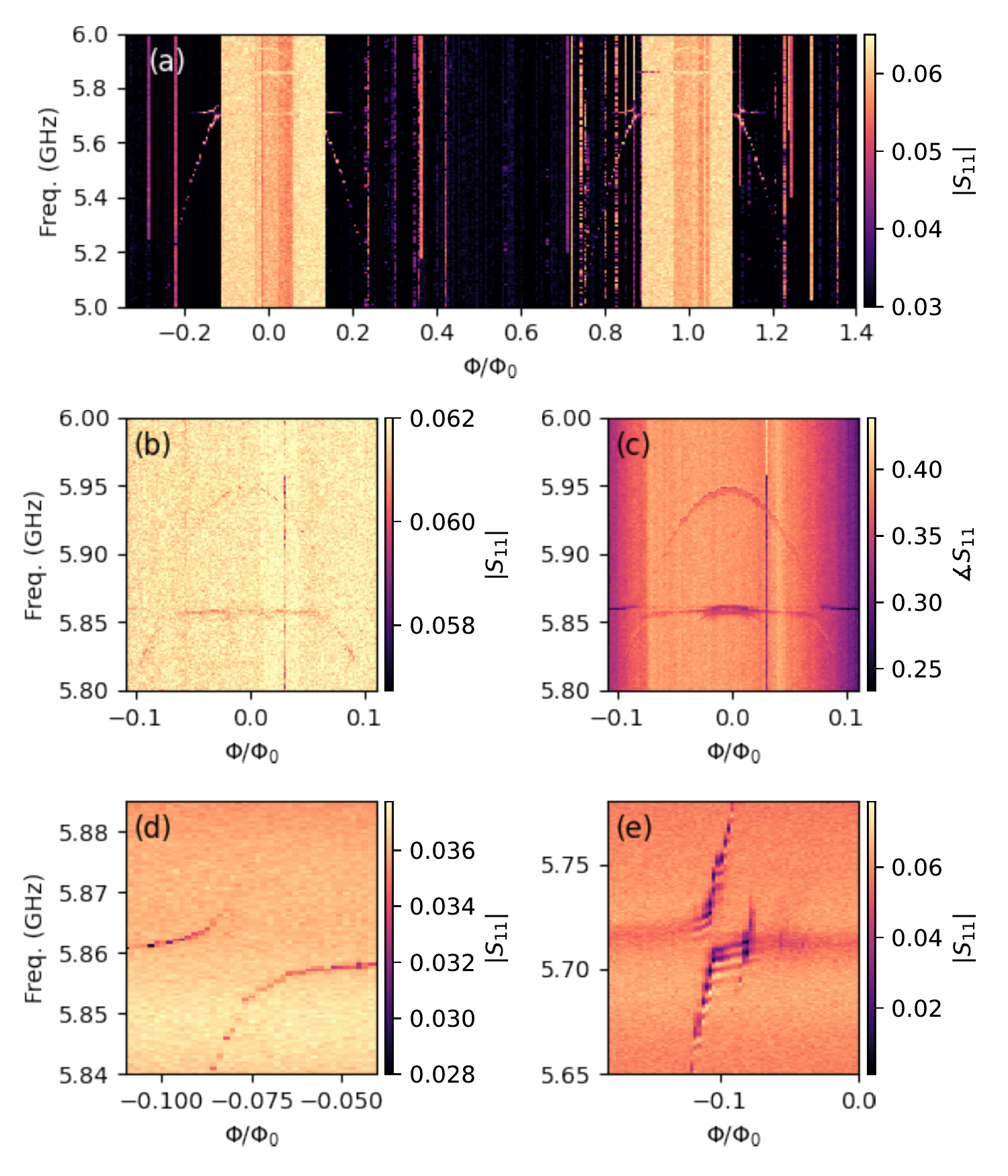}
		\caption{\textbf{(a)} Two-tone spectroscopy of two periods of the transmon. Two resonator lines at 5.86 GHz and 5.715 GHz are also visible. \textbf{(b,c)} Amplitude and phase of two-tone spectroscopy near the $f_{ge}^\text{max} = 5.95$ GHz. Reservoir line at 5.86 GHz is visible. \textbf{(d)} Single-tone spectroscopy of the avoided crossing with the reservoir. \textbf{(e)} Single-tone spectroscopy of the avoided crossing with the auxiliary resonator.}\label{app_qubit}
	\end{figure}

	The transmon parameters were found using two-tone spectroscopy (\autoref{app_qubit}(a-c)) with an additional microwave source. Its maximal frequency is $f_{ge}^\text{max} = 5.95$ GHz, as one can see from \autoref{app_qubit}(b,c). 
	
	In \autoref{app_qubit}(c) when $\Phi_e/\Phi_0 = 0$, there is an additional pair of anticrossings with reservoir resonator, which is caused by coupling of two-photon $\ket{g} \leftrightarrow \ket{f}$ transition with reservoir. Based on these data, the frequency of that transition is equal to $f_{r}$ and find transmon anharmonicity $\alpha = -180$ MHz.
	
	To find the coupling of the qubit to the cavities, we use single-tone spectroscopy (\autoref{app_qubit}(d,e)). From \autoref{app_qubit}(d) we estimate that coupling with reservoir resonator is $g_{r}/2\pi = \Delta/2 = 11$ MHz, where $\Delta$ is the splitting of the anticrossing. Similarly, the coupling with the auxiliary cavity from \autoref{app_qubit}(e): $g_{a}/2\pi = 15.5$ MHz. Periodic spectral features of unknown nature are noticeable in that avoided crossing.
	
	\section{Photon number calibration}\label{app:photon_cal}
	
	Steady-state population $N_\text{ss}$ in reservoir resonator can be calculated from emmitted power, measured by spectrum analyzer, by the following equation:
	
	\begin{equation}\label{eq:Nss}
	    N_\text{ss} = \frac{P_e}{\kappa_{r}^{e} h f_{r}},
	\end{equation}
	where $P_{e}$ is integral emitted power in Watts and
	\begin{equation}
	    P_e = P_\text{SA}/G,
	\end{equation}
	where $G$ is amplification in output line and $P_\text{SA}$ is the emission power measured by the signal analyzer. $G$ is the only unknown parameter and the task to calibrate $N_\text{ss}$ is reduced to amplification calibration.

	\begin{figure}[h!]
	    \vspace{0.5cm}
		\centering\includegraphics[width=0.8\textwidth]{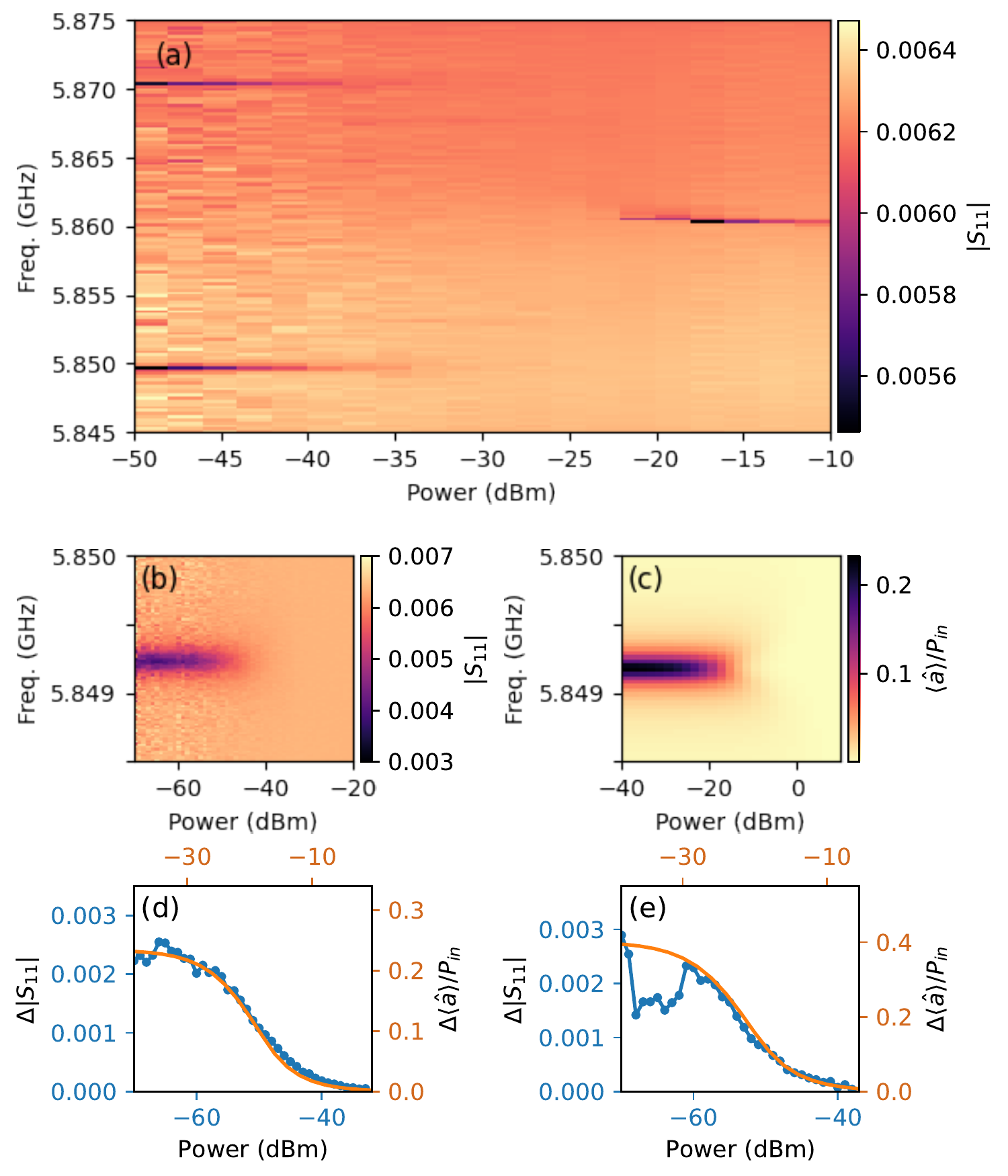}
		\caption{\textbf{(a)} Single-tone spectroscopy of qubit-reservoir anticrossing depending on VNA power. Lasing pumping is turned off. \textbf{(b)} Single-tone spectroscopy of bottom line of the anticrossing on (a) depending on VNA power. \textbf{(c)} Simulation of single-tone spectroscopy on (b): $\langle \hat{a} \rangle$ of reservoir depending on direct resonator pumping frequency and power. \textbf{(d)} Amplitude of Lorentz dip of $|S_{11}|$ from (b) (blue) and of Lorentz peak of $\langle \hat{a} \rangle$ from (c) (orange) depending on VNA power. \textbf{(e)} Same as (d), but for top line (analogous to (b) and (c) pictures are not shown)}\label{app_ph}
	\end{figure}

	To do that, we first need to measure the saturation of the transmon-reservoir anticrossing under the influence of increasing VNA power $P_\text{VNA}$ and compare it with the simulation. In \autoref{app_ph}(a), one can observe a smooth transition from a pair of spectral lines to a single line. Then, the top and bottom lines are measured separately with higher resolution (the bottom line is in \autoref{app_ph}(b), the top line is not shown). The next step is a numerical simulation of \autoref{app_ph}(b), which is shown in \autoref{app_ph}(c); we plot the magnitude of the mean value of the annihilation operator $|\langle \hat{a_r} \rangle|$ of the reservoir, which is proportional to $|S_{11}|$. Next, we fit vertical slices of \autoref{app_ph}(b) and (c) with Lorentzian curves and plot the amplitude of the dip (peak) depending on power both for experiment and simulation. We also repeat this procedure for top line of the avoided crossing. The final results for two lines are in \autoref{app_ph}(d) and (e). By comparing the values on the $x$-axes of the experimental and simulated plots, we find a difference of 31.5 dBm (this value is the same for both lines). The driving amplitude in our simulation depends not only on the absolute power at the cavity coupling port $P_\text{c.p.}$, but also on the external quality factor of the cavity, so this difference is not enough for calibration. We make another simulation with turned-off coupling between the cavity and the transmon, and then compare the result to a common formula for a bare cavity photon population,
	\begin{equation}\label{eq:Nph}
	    N_\text{ph} = \frac{4\kappa_{r}^{e}}{h f_{r} \kappa_{r}^2} P_\text{c.p.}
	\end{equation}
	In this additional simulation, for -3.5 dBm we obtain $N_\text{ph} = 33.5$. Then we know that the same population is reached in experiment for $P_\text{VNA} = -3.5-31.5=-35$ dBm. Using \eqref{eq:Nph}, we find that line attenuation $A = P_\text{VNA}/P_\text{c.p.} = 102.8$. We check that this value adds up well from the approximately 70 dB in the attenuators inside the refrigerator, 20 dB in the directional coupler that was used for two-tone spectroscopy and 10 dB of loss in the cables.
	
	Consequently, if the input power is -40 dBm, then $P_\text{c.p.}=-142.8$ dBm. Knowing this fact, we can find the amplification $G$ of output line. To do it, we set input power to -40 dBm, frequency to 5.9 GHz (to be not very far from working frequency) and measure the reflected power by spectrum analyzer, assuming that reflected power outside the resonance is almost equal to $P_\text{c.p.}$. The measured power is 78.8 dBm, which means that $G = 78.8-(-142.8) = 64$ dBm. Now we can use \autoref{eq:Nss} to calculate $N_\text{ss}$ for any output power.

	\section{Simplified analytical model with detuning}
	
	Since in the experimental configuration the $ef$ transition of the qubit is detuned by 35 MHz from the $\ket{0}_a\rightarrow\ket{1}_a$ transition, the calculation in Sec. IV of the theoretical proposal \cite{sokolova2021single} has to be modified. Precisely, we again need to solve the GKSL master equation, but now with a non-zero detuning included into the Hamiltonian:
	\begin{equation}\label{eq:jc}
	    \hat H_\text{t-a}/\hbar = g_a(ba^\dag_a + b^\dag a_a) + \delta b^\dag b,
	\end{equation}
	where $a_a$ is truncated to two lowest levels and $b$ to two levels $e$ and $f$ (meaning that the effective coupling constant should be changed to $g = g_a\sqrt{2}$). For this calculation, the dissipative part in the master equation contains only the collapse operator $\sqrt{\kappa_a} a_a$. We find that solving the resulting ODE system is easier in the real-number representation of the density matrix. If the corresponding transformation is done as $\rho_r = \mathcal{T}\rho$ with some superoperator $\mathcal{T}$, then the GKSL equation is transformed to
	\begin{equation}
	    \dot\rho_r = \mathcal{L}_r\rho_r, 
	\end{equation}
	where $\mathcal{L}_r=\mathcal{T}\mathcal{L}\mathcal{T}^{-1}$ is also real. Finding the effective decay rate $\gamma_{ef}^\text{eff}$ of the $ef$-transition ($\kappa_a^\text{eff}$ in the notation of \cite{sokolova2021single}) is now equivalent to finding the slowest-decaying non-oscillating exponent in the solution. The relevant eigenvalue of $\mathcal{L}$ turns out to be
	\begin{equation}\label{eq:kappa_eff}
	\begin{gathered}
	    \lambda = -\gamma_{ef}^\text{eff} =  -\frac{\kappa_{a}}{2}+\sqrt {\gamma^2+2 g_\delta^2},\\
 \gamma^2 = \frac{{\kappa_{a}}^{2}}{8}-\frac{{\delta}^{2}}{2}-2\,g^{2},\\
 g_\delta^2 = \sqrt { \left[ \frac{\delta^{2}}{4}+ \left(g+\frac{\kappa_{a}}{4}
 \right) ^{2} \right]  \left[ \frac{\delta^{2}}{4}+ \left(g-\frac{\kappa_{a}}{4}
 \right) ^{2} \right]  }.
    \end{gathered}
	\end{equation}
	For the experimental parameters $g/2\pi = 21.9$ MHz, $\kappa=90$ $\mu s^{-1}$, $\delta/2\pi = 35$ MHz, we find $\gamma^2 \approx -61.1\cdot10^3\ \mu s^{-2}$, $g_\delta^2 \approx 30.9\cdot10^3\ \mu s^{-2}$ and $\gamma_{ef}^\text{eff} \approx 17\ \mu s^{-1}$. One can compare this value with the resonant under-damped case where $\gamma_{ef}^\text{eff} = \kappa_a/2 = 45\ \mu s^{-1}$ \cite{sokolova2021single}. 
	
\begin{figure}
        \centering
        \includegraphics[width=.49\textwidth]{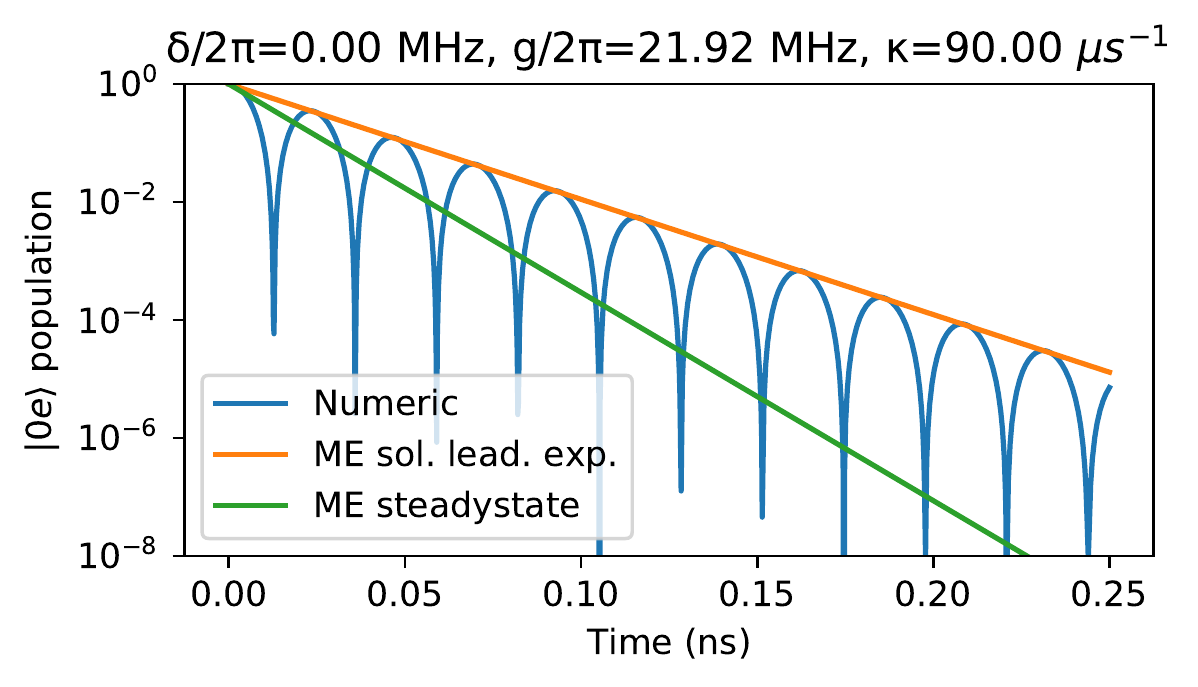}
    \includegraphics[width=0.49\textwidth]{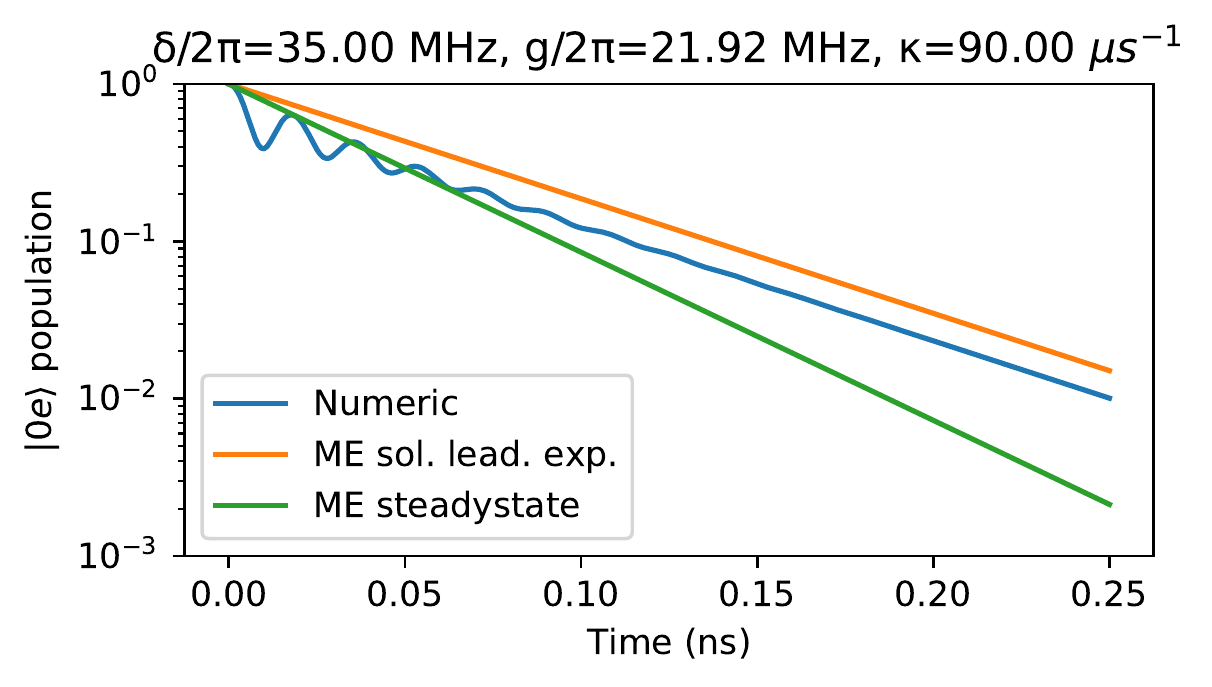}
        \caption{Comparison between the numerical solution (blue curve) and the two analytic solutions \eqref{eq:kappa_eff}, \eqref{eq:kappa_eff_2} (orange and green curves, respectively).}
        \label{fig:comparison_kappa_eff}
    \end{figure}

	Another way to calculate $\gamma_{ef}^\text{eff}$ is to find the steady state solution of the master equation with an addition of a weak incoherent excitation $\ket{e}\rightarrow \ket{f}$ at rate $\tilde\Gamma$. Lifting the two-level approximation for the cavity, we rewrite \eqref{eq:jc} as the Jaynes-Cummings model with detuning
	\begin{equation}
	    \hat H_\text{t-a}/\hbar = g(\sigma^-_{ef} a^\dag_a + \sigma^+_{ef} a_a) + \delta \sigma^+_{ef}\sigma^-_{ef},
	\end{equation}
	where $g=\sqrt{2}g_a$, and insert the collapse operators $\sqrt{\kappa_a} a_a$ and $\sqrt{\tilde\Gamma} \sigma^+_{ef}$ in the GKSL equation. In matrix form for the vector of the steadystate expectation values $\mathbf{x}_\text{ss} = \left[\left<a_a^\dag a_a\right>_\text{ss}\right.$, $\left<\sigma^+_{ef} \sigma^-_{ef} \right>_\text{ss},$ $\left<\sigma^+_{ef} a_a + \sigma^-_{ef} a^\dag_a \right>_\text{ss}$, $\left.\left<\sigma^+_{ef} a_a - \sigma^-_{ef} a^\dag_a \right>_\text{ss}\right]^T$ it reads:
	\begin{equation}
	    \left[ \begin {array}{cccc} -\kappa_{a}&0&0&ig
\\ \noalign{\medskip}0&-\tilde\Gamma&0&-ig
\\ \noalign{\medskip}0&0&-\kappa_{a}/2-\tilde \Gamma/2&i\delta
\\ \noalign{\medskip}2\,ig - 4\,ig\alpha&-2\,ig&i\delta&-\kappa_{a}/2-\tilde\Gamma/2\end {array} \right] \mathbf{x}_\text{ss} = \left[ \begin {array}{c} 0\\ \noalign{\medskip}-\tilde \Gamma
\\ \noalign{\medskip}0\\ \noalign{\medskip}0\end {array} \right] .
	\end{equation}
    Here, to linearize the system we replace the term $-4ig\left<\sigma^+_{ef}\sigma^-_{ef} a_a^\dag a_a \right>_\text{ss}$ in the last equation by $-4ig\left<a_a^\dag a_a\right>_\text{ss} \alpha$, where $\alpha=\left<\sigma^+_{ef}\sigma^-_{ef}\right>_\text{ss}$, neglecting the correlation between the qubit and resonator populations. In the steady state, the balance condition $\gamma_{ef}^\text{eff} \left<\sigma^+_{ef}\sigma^-_{ef}\right>_\text{ss} = \kappa_a \left<a_a^\dag a_a\right>_\text{ss}$ is satisfied, meaning that the rate of qubit population leakage must be equal to the rate of the cavity population leakage, since there is no other dissipation channel for the former. In the limit of $\tilde\Gamma\rightarrow0$ and $\alpha \rightarrow 0$ we obtain
    \begin{equation}\label{eq:kappa_eff_2}
    \gamma_{ef}^\text{eff} = \frac{\kappa_a \left<a_a^\dag a_a\right>_\text{ss}}{\left<\sigma^+_{ef}\sigma^-_{ef}\right>_\text{ss}} \approx {\frac {4\,{g}^{2}\kappa_{a}}{4\,{\delta}^{2}+4\,{g}^{2}+\kappa_{a}^2}}.
    \end{equation}
    This expression has a more compact form than \eqref{eq:kappa_eff}. We check that $\alpha \ll 1$ if $\tilde \Gamma \lesssim 5\ \mu s^{-1}$ by solving the quadratic self-consistency equation $\alpha=\left<\sigma^+_{ef}\sigma^-_{ef}\right>_\text{ss}$ for $\alpha$. For the experimental parameters, from \eqref{eq:kappa_eff_2} $\gamma_{ef}^\text{eff} \approx 25\ \mu s^{-1}$. The comparison between the numerical solution, \eqref{eq:kappa_eff}, and \eqref{eq:kappa_eff_2} is shown in \autoref{fig:comparison_kappa_eff} in log-scale; one can see that \eqref{eq:kappa_eff_2} predicts higher decay rate compared to the time-dependent solution though being closer to the numerical result in terms of RMS distance.
    
	Next, we estimate the effective pumping rate from $\ket{g}\rightarrow\ket{e}$ when the two-photon $ef$ transition Rabi frequency is $\Omega_\text{2ph}$. Compared to the proposal, the experimental value of $\gamma_{ef}^\text{eff}$ is now closer to the estimated $\Omega_\text{2ph}$ so that the strong-coupling case $\gamma_{ef}^\text{eff} < 2 \Omega_\text{2ph}$ may be reached. That means that the pumping rate $\Gamma = \gamma_{ef}^\text{eff}/2 \approx 8.5\ \mu s^{-1}$ by \eqref{eq:kappa_eff} or even $12.5\ \mu s^{-1}$ by \eqref{eq:kappa_eff_2}, compared to $5.3\ \mu s^{-1}$ calculated before for the optimal parameters in the resonant case \cite{sokolova2021single}. One can see that non-zero detuning does not deteriorate the effective pumping rate if compensated by the strong coupling regime of the two-photon pump. 
	
	Finally, we estimate the resulting expected population of the reservoir $N_\text{ss}$. The strong coupling condition $4\sqrt{N}g_r > N\kappa_r$ is met for $\kappa_r = 0.69\ \mu s^{-1},\ g_r/2\pi = 11$ MHz for $N<360$, so the expected population for the more modest $\Gamma$ value predicted by \eqref{eq:kappa_eff} is
	\begin{equation}
	    N_\text{ss} = 2\frac{\Gamma}{\kappa_r} \approx 24.
	\end{equation}
	
	\section{Numerical simulation}\label{app:simulation}

	\begin{figure}
		\centering\includegraphics[width=0.6\textwidth]{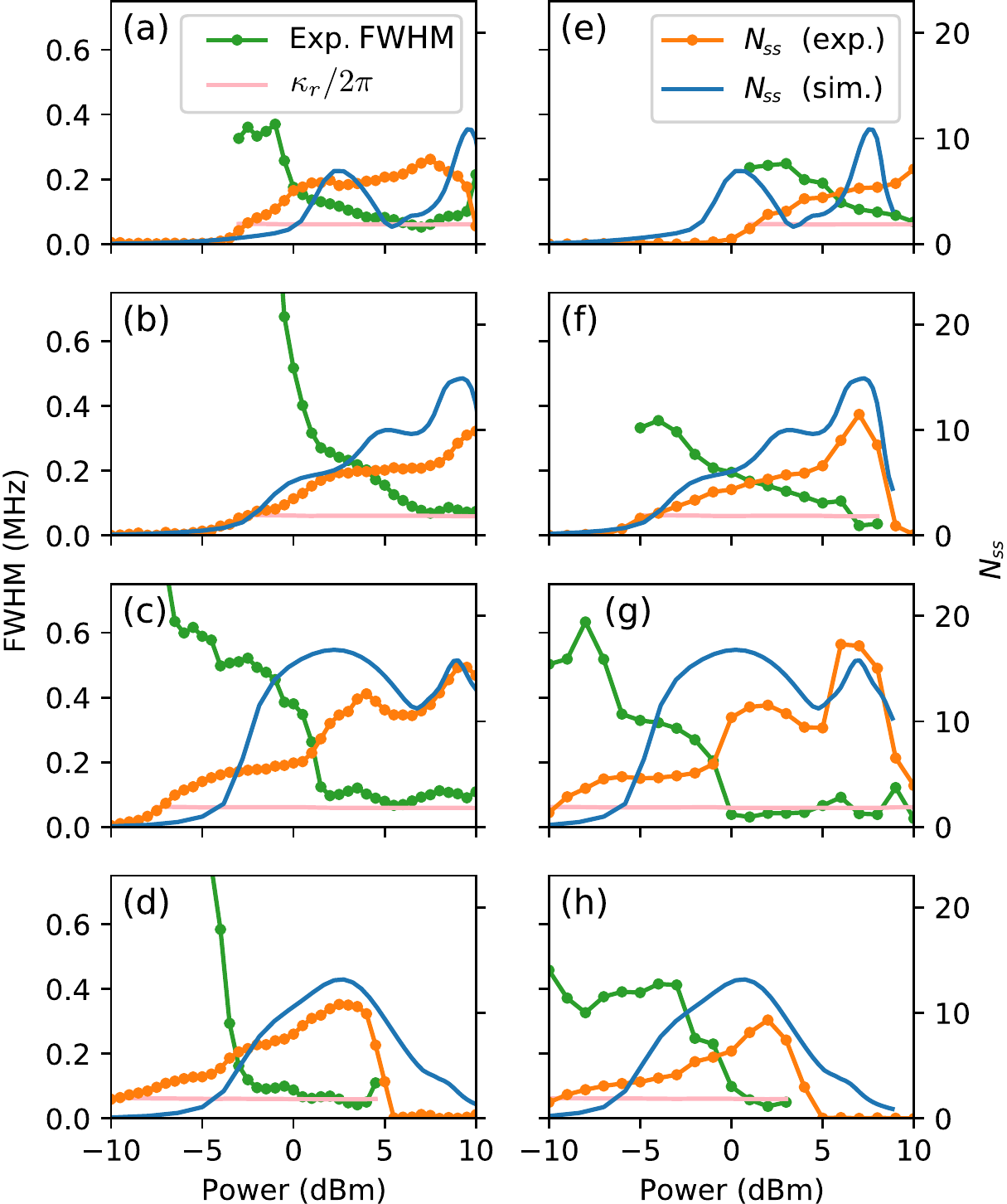}
		\caption{\textbf{(a-d)} First cooldown: number of pumped photons (experiment and simulation) and FWHM (experiment and pure reservoir resonator) for different pumping frequencies $\omega_{drive}$: (a,e) 5.7775 GHz, (b,f) 5.7825 GHz, (c,g) 5.7875 GHz~-- exact two-photon frequency, (d,h) 5.7925 GHz. \textbf{(e-h)} Second cooldown, parameters same as in the first cooldown.}\label{app_sim}
	\end{figure}
	
	Besides the analytical calculation, we use a full numerical model to check our results in a similar manner to the theoretical proposal \cite{sokolova2021single}. The simulation was performed using Python package QuTiP \cite{qutip}. The parameters used in simulation were taken from the experiment (see Table 1 of the main text): $f_{r} = 5.860$ GHz, $\kappa_{r} = 0.69$ $\mu \text{s}^{-1}$, $f_{a} = 5.715$ GHz, $\kappa_{a} = 90$ $\mu \text{s}^{-1}$, $g_{r}/2\pi = 11$ MHz, $g_{a}/2\pi = 15.5$ MHz. We assume that the transmon frequency is set to $f_{ge} = 5.863$ GHz, and the two-photon pumping frequency to be $f_\text{p}/2\pi = \frac{f_{r} + f_{a}}{2} = 5.7875$ GHz, which was used in the experiment, unless otherwise stated.
	
	The results of $N_\text{ss}$ depending on the pump power for different $f_\text{p}$ are shown on \autoref{app_sim} in comparison with the experiment for the two cooldowns. One can see that the experimental behavior reproduces itself with good accuracy from cooldown to cooldown. We note that numerical $N_\text{ss}$ agrees much better with the experimental results than the simple model prediction from the previous section, even replicating the self-quenching effects at high pumping powers.

	\section{Sample fabrication}
	
	A transmon qubit in this study is composed of two $\text{Al}/\text{Al}\text{O}_x/\text{Al}$ Josephson junctions (asymmetric SQUID) and a shunting capacitor. The JJs are patterned by electron lithography. The aluminum film is evaporated in HV e-beam system and the JJs are formed by double-angle Dolan bridge technique \cite{dolan1977, liechao2012} followed by lift-off. To build a ground plane, $100\,\text{nm}$ aluminum film is evaporated on a silicon substrate and etched in $\text{Cl}_2$ plasma after optical lithography. To improve the interface between the substrate and the ground metal we remove the native oxide by dipping the silicon wafer in piranha solution and buffered HF \cite{bruno2015, kalacheva2020} before aluminum evaporation. To obtain a good galvanic contact between the transmon qubit and the ground plane, it is further shorted by additional aluminum bandage layer \cite{osman2021}. We also implement aluminum air-bridges on the top of the coplanar transmission lines \cite{chen2014}. 
	
	\bibliography{main}